\documentclass[
reprint,
nofootinbib,
 amsmath,amssymb,
 aps,
 preprintnumbers,
]{revtex4-2}

\usepackage{graphicx}
\usepackage{dcolumn}
\usepackage{bm}

\usepackage{amssymb}
\usepackage{amsmath}
\usepackage{xcolor}
\usepackage{bbm}
\newcommand{\be}{\begin{equation}}
\newcommand{\ee}{\end{equation}}
\usepackage{hyperref}

\begin{document}

\preprint{HIP-2026-12/TH}

\title{
Landau-Zener formula and resonant axion conversion in neutron star magnetospheres
}

\author{Matti Heikinheimo} 
\email{matti.heikinheimo@helsinki.fi}
\affiliation{Department of Physics, University of Helsinki, 
                      P.O.Box 64, FI-00014 University of Helsinki, Finland}
\affiliation{Helsinki Institute of Physics, 
                      P.O.Box 64, FI-00014 University of Helsinki, Finland}

\author{Topi Sirkiä}
\email{topi.sirkia@helsinki.fi}
\affiliation{Department of Physics, University of Helsinki, 
                      P.O.Box 64, FI-00014 University of Helsinki, Finland}
\affiliation{Helsinki Institute of Physics, 
                      P.O.Box 64, FI-00014 University of Helsinki, Finland}
                      
\author{Kimmo Tuominen}
\email{kimmo.i.tuominen@helsinki.fi}
\affiliation{Department of Physics, University of Helsinki, 
                      P.O.Box 64, FI-00014 University of Helsinki, Finland}
\affiliation{Helsinki Institute of Physics, 
                      P.O.Box 64, FI-00014 University of Helsinki, Finland}

\date{\today}

\begin{abstract}
We investigate the Landau-Zener description of resonant axion-photon conversion in neutron star magnetospheres.
We find that this picture often fails for axion conversions to millimeter-to-optical band photons due to the characteristic resonance width exceeding the size of the conversion region. 
This comparison of scales yields a simple criterion for evaluating the validity of the Landau-Zener formula. 
We verify this criterion numerically, and show that when invalid, the Landau-Zener conversion probability may significantly deviate from the numerical result. In light of these findings, we revise constraints on axions from neutron star optical-band polarization searches.

\end{abstract}

\maketitle


\section{Introduction}

The quantum chromodynamics (QCD) axion and axion-like particles (ALPs) have emerged as one of the most promising candidates for physics beyond the Standard Model (SM). The QCD axion originally arose in the Peccei-Quinn solution to the strong CP problem~\cite{Peccei:1977hh,Peccei:1977ur,Weinberg:1977ma}, but it was soon noticed that it may also naturally comprise dark matter (DM)~\cite{Abbott:1982af,Preskill:1982cy,Dine:1982ah} and is naturally embedded in frameworks which are designed to solve other problems present in the SM~\cite{Albertus:2026fbe}. ALPs on the other hand generically arise in models with broken U(1) symmetries~\cite{Chikashige:1980ui,Davidson:1981zd,Wilczek:1982rv}, and in string theory~\cite{Witten:1984dg}. Hereafter we refer to both the QCD axion and ALPs collectively as just axions.

A wide range of laboratory and astrophysical searches exploit the axion-photon coupling~\cite{Sikivie:1983ip,Raffelt:1987im,DiLuzio:2020wdo,Sikivie:2020zpn,Graham:2015ouw,OHare:2024nmr}. Laboratory experiments include broadband searches such as helioscopes and light-shining-through-wall experiments, and resonant searches such as haloscopes and LC circuits~\cite{Bahre:2013ywa,CAST:2017uph,ADMX:2009iij,ADMX:2018gho,ADMX:2019uok,ADMX:2021nhd,Crisosto:2019fcj,DMRadio:2022pkf}. 
The strongest astrophysical constraints come from leveraging the strong magnetic fields of compact objects such as neutron stars (NSs) and magnetic white dwarfs (MWDs)~\cite{Carenza:2024ehj}. While such constraints are subject to larger astrophysical uncertainties, they generally do not require axions to constitute all of the dark matter.

Axion-photon conversion in magnetized plasmas may occur resonantly, when the axion mass matches the in-medium photon mass, or non-resonantly through mixing in strong magnetic fields. Resonant conversion in NS magnetospheres has been extensively studied, particularly in the context of radio searches for axion dark matter~\cite{Huang:2018lxq,Hook:2018iia,Leroy:2019ghm}. More recently, interest has increased in polarization-based searches at optical to X-ray frequencies~\cite{Benabou:2025jcv,Song:2024rru,Gau:2023rct}. At these higher frequencies, however, vacuum birefringence suppresses the conventional non-resonant conversion in the strong magnetic fields of neutron stars, making the comparatively weaker fields of MWDs more favorable for non-resonant conversion~\cite{Lai:2006af,Gill:2011yp,Dessert:2022yqq}. Consequently, polarization measurements of MWDs currently provide leading constraints on the axion-photon coupling for $m_a \lesssim 2\times10^{-7},\mathrm{eV}$~\cite{Benabou:2025jcv}, with complementary limits obtained from MWD light-curve observations~\cite{Tian:2025lkp}.

Interestingly, vacuum polarization also gives rise to an additional resonance that is enhanced, rather than suppressed, by the strong magnetic fields of neutron stars~\cite{Lai:2006af}.
Ref.~\cite{Bondarenko:2022ngb} showed that this resonance leads to distortions in the millimeter-band of spectral measurements, with potentially strong constraining power.
Using existing neutron star optical-band polarization measurements, Ref.~\cite{Song:2024rru} derived strong constraints on the axion-photon coupling for axion masses below $10^{-6}-10^{-3}$ eV, where the range of this upper bound depends on the plasma pair multiplicity. 

Often the analyses described above, use the Landau-Zener (LZ) approximation to compute the resonant axion-photon conversion probability. Hence, to model the signals and derive 
the related constraints accurately, it is important to establish the regime of validity of this approximation and understand when a fully numerical treatment is required. In this work we investigate the validity of the Landau-Zener approximation for resonant axion-photon conversion in stellar magnetospheres. We first demonstrate that, in the low adiabaticity limit, the Landau-Zener formula is equivalent to an all-orders perturbative stationary phase solution of the photon propagation equation. We then derive analytic conditions governing the existence and width of resonances, and show that for resonances producing optical and higher-frequency photons the assumptions underlying the Landau-Zener approximation are frequently violated because the characteristic resonance width becomes comparable to, or exceeds, the size of the conversion region. In this regime the Landau-Zener approximation systematically overestimates the conversion probability, as we explicitly demonstrate through numerical solutions of the propagation equations. Finally, we revisit polarization constraints on axions using optical frequency observations of pulsar B0656+14 and magnetar 4U 0142+61.

The paper is organized as follows: In Section~and~\ref{sec:resonances} we review the standard theory of axion-photon conversion, and connect the stationary phase approximation  result to the LZ result. In Sec~\ref{sec:anatomy} we discuss the validity of the LZ approximation in terms of the resonance location and width. 
In Sec.~\ref{sec:polarization} we apply our results to linear polarization degree measurements and present updated constraints on axions. Finally in Sec.~\ref{sec:conclusions} we conclude and discuss future prospects in polarization searches.

\section{Resonant axion-photon conversions}
\label{sec:resonances}

\subsection{Standard Theory}\label{sec:theory}
We begin by reviewing the standard approach to resonant axion-photon conversions in strong magnetic fields. The physics is encoded in axion-photon Lagrangian augmented with the Euler-Heisenberg terms,
\begin{align}
    \mathcal{L}&=-\frac{1}{4}F^2+\frac{1}{2}(\partial_\mu a)^2-\frac{1}{2}m_a^2 a^2 -\frac{g_{a\gamma}}{4}aF\tilde{F} \notag \\ &+\frac{\alpha^2}{90m_e^4}\big(F^2\big)^2+\frac{7\alpha^2}{360m_e^4}\big(F\tilde{F}\big)^2.
    \label{eq:lagrangian}
\end{align}
This effective Lagrangian is valid at photon energies below the electron mass. 

In the relativistic limit and keeping only the leading terms in the background magnetic field, $\propto B^2$, the linearized field equations can be cast in the form of a Schrödinger equation 
\be
(i \partial_r+{\cal{H}}(r)){\cal A}(r)=0, \label{eq:schrödinger}
\ee
where ${\cal A}=(a,A_{\|},A_\perp)^T$ and the Hamiltonian is
\be 
{\cal H}=\begin{pmatrix}
\Delta_a & \Delta_{\mathrm{M}} & 0 \\
\Delta_{\mathrm{M}} & \Delta_{\|}+\Delta_{\mathrm{pl}} & 0 \\
0 & 0 & \Delta_{\perp}
\end{pmatrix}.
\label{eq:Hamiltonian}
\ee
The matrix elements are given by
\begin{align}
\Delta_{\mathrm{M}} & \equiv\frac{1}{2} g_{a \gamma} B \sin \theta, \quad \Delta_a \equiv-\frac{m_a^2}{2 \omega}, \quad \Delta_{\mathrm{pl}} \equiv-\frac{\omega_{\mathrm{pl}}^2}{2 \omega} \notag \\
\Delta_{\|} &  \equiv\frac{7}{2} \omega \xi \sin ^2 \theta, \quad \Delta_{\perp} \equiv2 \omega \xi \sin ^2 \theta ,
\end{align}
where $\omega$ is the photon energy, $\xi=\alpha/(45\pi)(B/B_{\mathrm{c}})^2$ with $B_{\mathrm{c}}=m_e^2/e\sim 4.41 \times 10^{13}\,\, \mathrm{ G}$ the critical field strength and $\theta$ is defined as the angle between the external magnetic field and the direction of photon propagation (see App.~\ref{app:magnetosphere}).

As seen from Eq.~\eqref{eq:Hamiltonian}, the axion couples only to the $A_\parallel$-polarization. This asymmetry allows axion conversion to polarize initially unpolarized light, such as thermal emission of stars. Conversion between photons and axions is governed by the mixing angle obtained by diagonalizing the upper block of the Hamiltonian in Eq.~\eqref{eq:Hamiltonian}:
\begin{align}
\tan 2 \theta_\mathrm{mix}=\frac{2 \Delta_{\mathrm{M}}}{\Delta_{\|}+\Delta_{\mathrm{pl}}-\Delta_a}.\label{eq:mixingangle}
\end{align}

In very strong magnetic fields the mixing angle scales as $\sim1/B$, and axion-photon conversion becomes suppressed by vacuum birefringence effects. As such, 
conversion in magnetic white dwarves (MWDs) is often more efficient 
and has been used to set stringent constraints 
on the axion-photon coupling~\cite{Dessert:2022yqq,Benabou:2025jcv}. 
However, this suppression is not present at resonance, where the denominator of the mixing angle of Eq.~\eqref{eq:mixingangle} is close to zero.

To compute the asymptotic axion photon conversion probability in the presence of a resonance, 
it is common practice~\cite{Battye:2019aco,Carenza:2023nck} to apply the Landau-Zener (LZ) formula~\cite{Landau:1932vnv,Zener:1932ws} 
\begin{align}
    P_\mathrm{LZ}&=1-e^{-2\pi\gamma_{\mathrm{res}}}
    \label{eq:LZformulabasic}
\end{align}
where the adiabaticity factor $\gamma_{\rm{res}}$ 
describes the strength of conversion. It is generally defined in terms of parameters in the Hamiltonian of Eq.~\eqref{eq:Hamiltonian} as 
\begin{align}
    \gamma_{\text{res}} &\equiv \frac{\Delta_\mathrm{M}^2}{|\partial_r(\Delta_{\parallel}+\Delta_{\text{pl}}-\Delta_a)|}\Big\vert_{r_{\mathrm{res}}} 
    \label{eq:adiabaticityfactor}.
\end{align}
The resonance radius $r_{\text{res}}$ is a
root of the denominator in the mixing angle of Eq.~\eqref{eq:mixingangle}. 
 
For axion-photon couplings not yet ruled out, the non-adiabatic conversion regime $\gamma_\mathrm{res}\ll 1$ is generally the relevant one. We will next solve the Schrödinger equation and obtain asymptotic solution equivalent to the LZ formula in the non-adiabatic limit and in the presence of a resonance. 

\subsection{Perturbative solution}

As the $A_\perp$-polarization decouples, the axion-photon mixing is governed by the two-state Hamiltonian
\be 
H=\begin{pmatrix} \Delta_a & \Delta_\mathrm{M} \\ \Delta_\mathrm{M} & \Delta_{\|}+\Delta_{\mathrm{pl}}\end{pmatrix}=H_0+H_\mathrm{M},
\label{eq:Hamiltonian2}
\ee 
where \mbox{$H_0={\mathrm{diag}}(\Delta_a,\Delta_{\|}+\Delta_{\mathrm{pl}})$} and $H_\mathrm{M}$ contains the off-diagonal elements. The Schrödinger equation with this Hamiltonian,
$
\left[i \partial_r+{H}(r)\right]{\psi(r)}=0$,
then determines the dynamics of the fields $\psi(r)=(a(r),A_{\|}(r))^T$.  

We obtain the perturbative solution in the interaction presentation, where the evolution operator is defined as  
\begin{align}
U(r) &=
\exp\left(i\int_R^r H_0(r')dr'\right) \notag\\
&=\begin{pmatrix} e^{i\Delta_a r} & 0 \\ 0 & e^{i\Delta_a r+i\int_R^r\Delta_{\text{tr}}dr'}\end{pmatrix},
\end{align}
with $\Delta_{\text{tr}}=\Delta_\|+\Delta_{\text{pl}}-\Delta_a$, and the states are defined by $\psi_{\text{int}}=U^\dagger\psi$ and the interaction Hamiltonian by 
\begin{align}
H_{\text{int}} &=U^\dagger H_I U \notag \\
&=\begin{pmatrix} 0 & \Delta_\mathrm{M}e^{i\int_R^r\Delta_{\text{tr}}(r')dr'} \\ 
 \Delta_\mathrm{M}e^{-i\int_R^r\Delta_{\text{tr}}(r')dr'} & 0 \end{pmatrix}.
\end{align} 
We solve the Schrödinger equation
\be 
-i\partial_r\psi_{\text{int}}=H_{\text{int}}\psi_{\text{int}} 
\ee 
iteratively as 
\be 
\psi_{\text{int}}^{(n+1)}(r)=i\int_R^rdr'H_{\text{int}}(r')\psi_{\text{int}}^{(n)}(r'),
\ee 
starting with the initial condition 
$\psi_{\text{int}}^{(R)}(r)=\psi(R)$.
Explicitly:
\begin{align}
\begin{pmatrix}
    a^{(n)}(r) \\ 
    A_\parallel^{(n)}(r)
\end{pmatrix}
=
&(i)^n
\int_{R_n}\prod_{i=1}^{n}\Delta_\mathrm{M}(r_i)dr_i
\begin{pmatrix} e^{i\phi}\psi_1(R) \\
e^{-i\phi}\psi_2(R)\end{pmatrix},
\end{align} 
where $R_n$ is a simplex defined by
\be 
r>r_1>r_2>\dots>r_{n}>R,
\ee 
and the phase factor is defined as 
\be 
\phi\equiv\phi(r_1,r_2,\dots,r_n)
=\sum_{i=1}^n\int_R^{r_i}\Delta_{\text{tr}}(r')dr'.
\ee
Finally, the initial values are given by $\psi_1(R)=a(R)$ and $\psi_2(R)=A_{\|}(R)$ for even $n$ and vice versa for odd $n$.

Note that the initial conditions $\psi(R)=(1,0)^T$ and $\psi(R)=(0,1)^T$ are particularly simple as then $a(r)$ and $A_{\|}(r)$ receive contributions only from either odd or even orders. Let us therefore focus on the initial condition $a(R)=0$ and $A_{\|}(R)=1$. Then $a(r)$ is given by the odd order terms and $A_{\|}(r)$ by the even order terms. From the expression above, we deduce that generally 
\begin{align} 
A_{\|}^{(2k)}(r) &=(i)^{2k}\int_{R_{2k}}
\prod_{i=1}^{2k}
dr_i\Delta_\mathrm{M}(r_i)e^{-i\phi(r_1,\dots,r_{2k})}A_\|(R).
\end{align}
The function $\phi$ in the exponential is in this case given by 
\be
\phi(r_1,\dots,r_{2k})=
\int_{r_2}^{r_1}\Delta_{\text{tr}}(r')dr'+\cdots+\int_{r_{2k}}^{r_{2k-1}}\Delta_{\text{tr}}(r')dr'.
\ee 
To evaluate this integral, we assume that it is dominated by a single 
resonance\footnote{For more resonances, as long as they do not overlap, the stationary phase treatment generalizes easily: the integral is evaluated for each resonance independently, and the total conversion probability is given by the sum of adiabaticities of each resonance.}, extend the integration to ${\mathbbm{R}}^{2k}$ and approximate the background magnetic field by its value 
at the position of the resonance. The stationary phase approximation
then gives the asymptotic solution as
\be
A_\|^{(2k)}=\frac{(-1)^k}{(2k)!}
\left(\frac{(2\pi\Delta^2_{\mathrm{M}})^k}
{(|{\text{det}}{\cal{Z}}|)^{1/2}}\right)e^{-i\Delta_{\text{tr}}+\frac{i\pi}{4}\text{sgn}({{\text{det}}\cal{Z}})}A_\|(R),
\label{eq:assolution}
\ee 
where ${\cal{Z}}$ is the Hessian, ${\cal{Z}}_{ij}=\partial_i\partial_j\phi(r_{\text{res}})$ with $i,j=1,\dots,2k$, 
and $\text{sgn}({\cal{Z}})$ is the difference of the number of its positive and negative eigenvalues. All quantities in Eq.~\eqref{eq:assolution} are to be understood as evaluated at $r=r_{\text{res}}$.
 
Adding up all these contributions yields 
\be 
A_\|=A_\|(R)e^{-i\Delta_{\text{tr}}(r_{\text{res}})}\cos\left(
\sqrt{2\pi\gamma_{\text{res}}}
\right),
\label{eq:aparallorder}
\ee 
where $\gamma_{\text{res}}=\Delta_\mathrm{M}^2(r_{\text{res}})/(\partial_r\Delta_{\text{tr}}(r_{\text{res}}))$ was identified as the adiabaticity factor of Eq.~\eqref{eq:adiabaticityfactor}.
A similar calculation for the odd terms gives 
\be 
a=A_\|(R)ie^{i\Delta_{\text{tr}}(r_{\text{res}})+i\pi/4}\sin\left(\sqrt{2\pi\gamma_{\text{res}}}
\right). \label{eq:aallorder}
\ee

From the above expressions we easily obtain the leading perturbative corrections to order ${\cal{O}}(\Delta_\mathrm{M}^3)$. Neglecting the unimportant overall phase, 
for the axion this is
\begin{align}
    a^{(1)}
&= \Delta_\mathrm{M}(r_{\rm{res}})L_{\mathrm{res}}, \label{eq:statphase}
\end{align}
where we have defined the characteristic resonance length scale 
\begin{align}
    L_{\mathrm{res}}\equiv \sqrt{\frac{2\pi}{|\phi''(r_{\rm{res}})|}}=\sqrt{\frac{2\pi}{|\partial_r(\Delta_{\rm pl}+\Delta_\parallel-\Delta_a)|\big|_{r_{\rm{res}}}}}. \label{lres}
\end{align}

The asymptotic axion-photon conversion probability is then 
\begin{align}
P_\mathrm{conv}(\infty)&=|a^{(1)}|^2=\Delta_\mathrm{M}^2(r_\mathrm{res})L_\mathrm{res}^2= 2\pi\gamma_\mathrm{res}. \label{eq:presamp}
\end{align}
This result exactly coincides with that of the LZ formula, Eq.~\eqref{eq:LZformulabasic}, 
in the non-adiabatic $\gamma_{\text{res}}\ll 1$ regime. Therefore, for our purposes the two are the same and we hereafter refer to the stationary phase result simply as the LZ result. For the photon, the leading perturbative correction arises at 
${\cal{O}}(\Delta_\mathrm{M}^2)$, and is given by 
\begin{align}
     A_\parallel^{(2)}&= -\frac{\Delta_\mathrm{M}^2}{2}\frac{2\pi}{|\partial_r(\Delta_\|+\Delta_{\text{pl}}-\Delta_a)|}\Big\vert_{r_{\text{res}}}
          &=-\pi \gamma_{\mathrm{res}} \label{eq:amp2ndorder}.
\end{align}

The conversion probability is now given as the complement of the photon survival probability $P_S$
\begin{align}
    P_\mathrm{conv}(r)&= 1-P_S (r)\equiv 1-\Big| 1+A^{(2)}_\parallel(r)\Big|^2 \notag \\ &\simeq -2 \Re \Big( A^{(2)}_\parallel(r)\Big). \label{Pconvr}
\end{align}
The equivalence with Eq.~\eqref{eq:presamp} follows from the 
unitarity of the Schrödinger evolution. With the initial condition $A_{\|}(R)=1$, at order ${\mathcal O}(\Delta_\mathrm{M}^2)$ this means
\be 
1=|a(r)|^2+|A_{\|}(r)|^2\simeq |a^{(1)}(r)|^2+1+2\Re \Big( A_{\|}^{(2)}(r) \Big) ,
\ee 
and hence $2\Re \Big(A_{\|}^{(2)}(r) \Big)\simeq -|a^{(1)}(r)|^2$. Asymptotically far, one obtains the relation
\begin{align}
    \Re \Big(A_{\|}^{(2)}(\infty )\Big)\simeq -\pi \gamma_\mathrm{res}.
\end{align}
Having established the relevant results for resonant conversion,  we will next describe in detail the conditions under which these results are consistent. 

\section{Resonance anatomy and validity of LZ formula}\label{sec:anatomy}

The LZ approximation relies on the resonance being localized enough that the background varies only weakly across the conversion region. Additionally the width should not overlap considerably with the star, and in cases of multiple resonances they should be effectively independent. Whether these assumptions hold is determined by three quantities: the resonance point $r_\mathrm{res}$, width $L_\mathrm{res}$ and adiabaticity $\gamma_\mathrm{res}$. In this section we derive analytic expressions for these quantities and use them to establish simple criteria for the validity of the LZ approximation. We find that for resonances relevant to optical and higher frequency polarization searches, these criteria are often violated.
 
As a concrete model for the magnetosphere, we adopt the Goldreich-Julian (GJ) model, and parametrize deviations from the GJ charge density with a plasma pair multiplicity factor $\kappa$ which typically takes values $\kappa\sim 1-10^4$~\cite{Timokhin:2018vdn}. In the figures of this section we have fixed the rotation phase to $\Omega t=0$ for simplicity.

\subsection{LZ validity conditions} \label{sec:lzvalcons}

We now describe the conditions mentioned earlier more quantitatively. We take the resonance width to be defined 
as the characteristic length $L_\mathrm{res}$ of the stationary phase prefactor in Eq.~\eqref{lres}. We then require that any resonance does not overlap with the star, $r_\mathrm{res}-\frac{1}{2}L_\mathrm{res}\gtrsim R$. This condition also automatically encodes the requirement that the resonance itself is outside the star. 

Additionally, we require the background magnetic field and plasma properties vary on scales larger than the resonance width, 
\be 
L_B(r_\mathrm{res})\equiv \Big|\frac{B(r_\text{res})}{\partial_r B(r_\text{res})}\Big|=r_\mathrm{res}/3 \gtrsim L_\mathrm{res},
\ee 
where the second equality holds for the GJ model. Combining the two conditions, we have
\begin{align}
    r_\mathrm{res}&\gtrsim \max (R+\frac{L_\mathrm{res}}{2},3L_\mathrm{res}) .\label{eq:rrescondition}
\end{align}
Hence, for $L_\mathrm{res}\lesssim \frac{2}{5}R$ the resonance width is the relevant constraint 
while for wider resonances the variation of the background becomes the determining factor.

In addition to these conditions for each resonance, in the multi-resonance case the resonances themselves must be well separated spatially, ie. their resonance lengths do not overlap, so that resulting stationary phase 
integrals can be performed independently:
\begin{align}
    |r_\mathrm{res}^+ -r_\mathrm{res}^-| \gtrsim \frac{L_\mathrm{res}^++L_\mathrm{res}^-}{2}. \label{eq:nooverlapcondition}
\end{align}
In this case the Landau-Zener survival probabilities of each resonance may simply be multiplied, giving a total adiabaticity $\gamma_\mathrm{eff}\equiv \gamma_\mathrm{res}^+ +\gamma_\mathrm{res}^-$.
Here we have denoted the two separate resonances with the indices $+,-$. To aid the evaluation of these conditions, we will now express the three resonance parameters in terms of the neutron star and axion parameters for each resonance.

The discussion here assumes a monotonic plasma profile, which may not be realized e.g. in turbulent magnetospheres. For an example of a more complicated multi-resonance scenario with non-monotonic plasma profile, see Ref.~\cite{Brahma:2023zcw}.

\subsection{Resonance parameters}

Let us start from the definition of the mixing angle of Eq.~\eqref{eq:mixingangle}, where the zeros of the denominator correspond to resonances. In literature~\cite{Song:2024rru,Long:2024qvd} the resonances are typically considered in the limits
\begin{align}
    |\Delta_{\mathrm{pl}}|&=|\Delta_a|\gg\Delta_\parallel,  \notag \\
    |\Delta_{\text{pl}}|&=\Delta_\parallel\gg |\Delta_a|. \label{eq:resonances}
\end{align}
We refer to the first case as the plasma resonance and the second one as the vacuum resonance. In order to proceed in full generality, we first describe the resonances in terms of the exact roots of the mixing angle denominator. The plasma and vacuum resonances of Eq.~\eqref{eq:resonances} will then be obtained as limits of the more general results. The distinction is important in the intermediate axion mass regime, where none of the $\Delta$-terms can be neglected. 

The resonances are roots of the sextic equation
\begin{align}
    Ar^{-6}+B r^{-3}+C&=0, \label{sextic}
\end{align}
where
\begin{align}
    A&\equiv \frac{7}{2}\frac{\alpha}{45\pi}\frac{B_0^2}{B_\mathrm{c}^2}R^6 \omega \sin^2{\theta} F(\theta_\mathrm{m}, \theta_\mathrm{o}, \Omega t)^2 =\Delta_\parallel r^6\notag, \\
    B&\equiv -\frac{8\pi^2 \alpha  \kappa}{\omega e m_e T}B_0 R^3 G(\theta_\mathrm{m}, \theta_\mathrm{o}, \Omega t)=\Delta_{\text{pl}}r^3, \notag \\
    C&\equiv \frac{m_a^2}{2\omega}=-\Delta_a,
\end{align}
where we have defined the auxiliary angular functions
\begin{align}
     F(\theta_\mathrm{m},\theta_\mathrm{o},\Omega t)&\equiv \sqrt{3{(\hat{m}\cdot\hat{r})}^2+1}, \notag \\
    G(\theta_\mathrm{m},\theta_\mathrm{o},\Omega t)&\equiv |3(\hat{m}\cdot\hat{r})\cos{\theta_\mathrm{o}} -\cos{\theta_\mathrm{m}}|,
\end{align}
with $\hat{m}\cdot\hat{r} \equiv \cos{\theta_\mathrm{m}}\cos{\theta_\mathrm{o}}+\sin{\theta_\mathrm{m}}\sin{\theta_\mathrm{o}}\cos{\Omega t}$. They encode the angular dependence in the magnetic field and its $z$-component, respectively. For small angles $\theta_\mathrm{m}$ and $\theta_\mathrm{o}$ we have, up to corrections at least quadratic in the angles, 
$F\simeq 2$ and $G\simeq 2$. 
For brevity of notation, we suppress the arguments of these functions in the following equations. 

The resonances are then generally given by
\begin{align}
    r_\pm \equiv {\Big( \frac{2 A}{-B\pm \sqrt{B^2-4AC}}\Big)}^{1/3}. \label{generalres}
\end{align}
We refer to the near resonance $r_+$ as vacuum-like and the farther $r_-$ as plasma-like to highlight what they become in appropriate limits. The vacuum resonance, if it exists, is always closer to the star than the plasma resonance.

The existence of resonances (real roots) is governed by the 
positivity of the discriminant, $B^2-4AC>0$. We note that the existence of real roots does not yet guarantee the resonance to be physical, ie. being outside of the star, for instance. 
From the discriminant positivity one can solve for the axion mass above which   the resonances merge and conversion becomes non-resonant:
\begin{align}
    m_a > \pi^2\sqrt{\frac{720}{7}} \frac{B_\mathrm{c}}{m_e}\frac{\kappa}{\omega T \sin{\theta}} \frac{G}{F}\simeq \sqrt{\frac{180 \pi^3}{7\alpha}}\frac{\kappa m_e}{\omega T}, \label{eq:masscondition}
\end{align}
where we took $G/F\simeq 1$ and $\sin\theta\simeq 1$ and used $B_{\text{c}}=m_e^2/e$. 
For instance, for reference values $T=8.689$ s and $\omega=0.1$ eV, one obtains $m_a> 1.25\times 10^{-7}$ eV.

The generalized adiabaticity factor is obtained 
analogously with Eq.~\eqref{eq:adiabaticityfactor}, 
by writing
\begin{align}
    \gamma_{\pm} &=\frac{\Delta_\mathrm{M}^2}{|\partial_r(A(r^{-3}-r_+^{-3})(r^{-3}-r_-^{-3})|}\Bigg\vert_{r_\pm}
    \notag \\
    &=\frac{D^2 r_\pm}{3}\frac{1}{|A-Cr_\pm^6|} \label{generalgamma},
\end{align}
where we have defined
\begin{align}
    D \equiv \frac{1}{2} g_{a\gamma } \sin \theta B_0 R^3 F=\Delta_\mathrm{M} r^3.
\end{align}
The resonance widths are given similarly as
\begin{align}
    L_\pm &= \sqrt{\frac{2\pi \gamma_\pm}{\Delta_\mathrm{M}^2(r_\pm)}}.
    \label{eq:widthsgeneral}
\end{align}

The plasma and vacuum resonances of Eq.~\eqref{eq:resonances} are obtained from Eq.~\eqref{sextic} setting $A\rightarrow 0$ and $C\rightarrow 0$, respectively: 
\begin{align}
    r_{\mathrm{res}}^{\mathrm{pl}}&=\left(-\frac{B}{C}\right)^{1/3}= {\Big(G\frac{16\pi^2\alpha}{em_e T}\frac{\kappa B_0}{m_a^2} \Big)}^{1/3} R, \notag \\
    r_{\mathrm{res}}^{\mathrm{vac}}&=\left(-\frac{A}{B}\right)^{1/3}= {\Big(\frac{F^2}{G}\frac{7em_e T}{720\pi^3}\frac{\omega^2 B_0}{\kappa B_{\mathrm{c}}^2}\sin^2\theta \Big)}^{1/3} R
    \label{resradii}.
\end{align}
The plasma resonance approaches the surface of the star as $r_{\text{res}}^{\text{pl}}\sim m_a^{-2/3}$, as larger plasma densities are required to match the axion mass. The location of the vacuum resonance is independent of $m_a$. On the other hand the vacuum resonance is inside the star for low frequencies, with location scaling as $r_{\text{res}}^{\text{vac}}\sim \omega^{2/3}$, which also explains why the vacuum resonance is not relevant for radio axion searches. The location of the plasma resonance is independent of $\omega$. Finally, we note that increased plasma multiplicity shifts the plasma resonance farther, while the vacuum resonance moves closer to the stellar surface.

The adiabaticities corresponding to the roots $r_{\mathrm{res}}^{\mathrm{pl}}$ and $r_{\mathrm{res}}^{\mathrm{vac}}$ are obtained 
from the first expression in the second line of Eq.~\eqref{generalgamma} by substituting $r_-=r_{\text{res}}^{\text{pl}},A=0$ for the plasma resonance and 
$r_+=r_{\text{res}}^{\text{vac}}$ for the vacuum resonance. These lead to
\begin{align}
    \gamma_{\mathrm{res}}^{\mathrm{pl}}&=\frac{\sin^2\theta}{ 48}\frac{F^2}{G^{5/3}}\omega R {\Big(\frac{g_{a\gamma}^6 B_0 m_a^4 (m_e T)^5}{2 e^5\pi^5\kappa^5} \Big)}^{1/3}, \notag \\ 
    \gamma_{\mathrm{res}}^{\mathrm{vac}}&=  g_{a\gamma}^2B_{\mathrm{c}}{\Big( \frac{F^2}{G} \frac{75}{98}\frac{\pi^3}{e^5 \kappa}\frac{m_e}{\omega} R^3 T B_0B_{\mathrm{c}}\sin^2{\theta}\Big)}^{1/3} \label{gammas}.
\end{align}
Notable about these scalings is that the conversion strengths of the two resonances scale oppositely with the photon frequency, $\gamma_{\mathrm{res}}^{\mathrm{pl}}\propto \omega$ and $\gamma_{\mathrm{res}}^{\mathrm{vac}}\propto \omega^{-1/3}$. This makes the vacuum resonance dominant in the millimeter-to-optical frequency band. We also note that both the plasma and vacuum resonances are weakened by large plasma pair multiplicities, although the damping is much stronger for the plasma resonance, $\gamma_{\mathrm{res}}^{\mathrm{pl}}\propto \kappa^{-5/3}$ and $\gamma_{\mathrm{res}}^{\mathrm{vac}}\propto \kappa^{-1/3}$.

The resonance widths of Eq.~\eqref{eq:widthsgeneral} for each of the resonances are given by
\begin{align}
    L_{\mathrm{res}}^{\text{pl}}&=
       \sqrt{\omega R}{\Big( G\frac{256e \pi^4 \kappa}{27 m_e T} \Big) }^{1/6}\left(\frac{B_0}{m_a^8}\right)^{1/6} , \notag \\ 
       L_{\text{res}}^{\text{vac}} &=
       \frac{7^{2/3}}{10^{2/3}}
       \frac{\sqrt{\omega R}\sin^{4/3}\theta}{12\cdot 3^{5/6}}
       {\Big( \frac{F^8}{G^7} 
       \frac{e}{\pi^{12}}{\frac{(m_e T)^7}{\kappa^7}} \Big) }^{1/6}\left(\frac{\omega^{8} B_0}{B_{\mathrm{c}}^8}\right)^{1/6} .
    \label{reswidth}
\end{align}

From Eq.~\eqref{reswidth} one observes that the resonances both grow wider with increasing photon frequency, $ L_{\mathrm{res}}^{\text{pl}}\propto \omega^{1/2}, L_{\mathrm{res}}^{\text{vac}}\propto \omega^{11/6}$. This eventually violates the conditions of Eqs.~\eqref{eq:rrescondition} and ~\eqref{eq:nooverlapcondition}. We note that the $\kappa$-scaling of the resonance width worsens this issue for the plasma resonance and alleviates it for the vacuum resonance. However, when both resonances are present and relevant, one needs to verify the validity conditions of Sec.~\ref{sec:lzvalcons} for both independently.

\subsection{Determining the validity of LZ result}

Let us now illustrate the use of these results. We consider the two neutron stars PSR B0656+14 and magnetar 4U 0142+61 as benchmark cases. Their stellar parameters obtained from Australia Telescope National Facility (ATNF) pulsar catalogue~\cite{Manchester_2005}, as well as observed 90\% confidence level linear polarization degrees (PD) in the optical band~\cite{Mignani:2015ufa,Wang:2015ppa} are listed in Table~\ref{tab:neutronstars}. In absence of precise knowledge of the magnetic and observation angles, we adopt the same values for both stars, $\theta_\mathrm{m}=30^\circ,\theta_\mathrm{o}=40^\circ$, fitted to radio-band measurements of PSR B0656+14~\cite{Kern:2003mj}.

\begin{table}[h!]
\centering
\begin{tabular}{|c|c|c|c|c|c|}
\hline
\textbf{Neutron Star} & PSR B0656+14 & 4U 0142+61 
\\
\hline
$B_0$ (G) & $4.7\times 10^{12}$  & $1.3\times 10^{14}$ 
\\
$T$ (s) & 0.384 & $8.689$ 
\\
$R$ (km) & $9.3$ & 16.1 
\\
$\Pi_\mathrm{L}$(\%)& $11.9 \,\pm \, 5.5$ & $<5.6$ 
\\
$\lambda$ (nm) & $555$ & $802$ \\
$\omega=2\pi/\lambda$ (eV)  & $2.23$ & $1.55$ 
\\
\hline
\end{tabular}
\caption{Stellar parameters $B_0$, $T$ and $R$ of the neutron stars considered in this work~\cite{Manchester_2005}. Also shown are the linear degree of polarization ($\Pi_\mathrm{L}$) measured and the corresponding observation frequency.
}
\label{tab:neutronstars}
\end{table}

\begin{figure*}[ht!]
    \begin{minipage}{0.4\linewidth}
       \centering
        \includegraphics[width=8 cm]{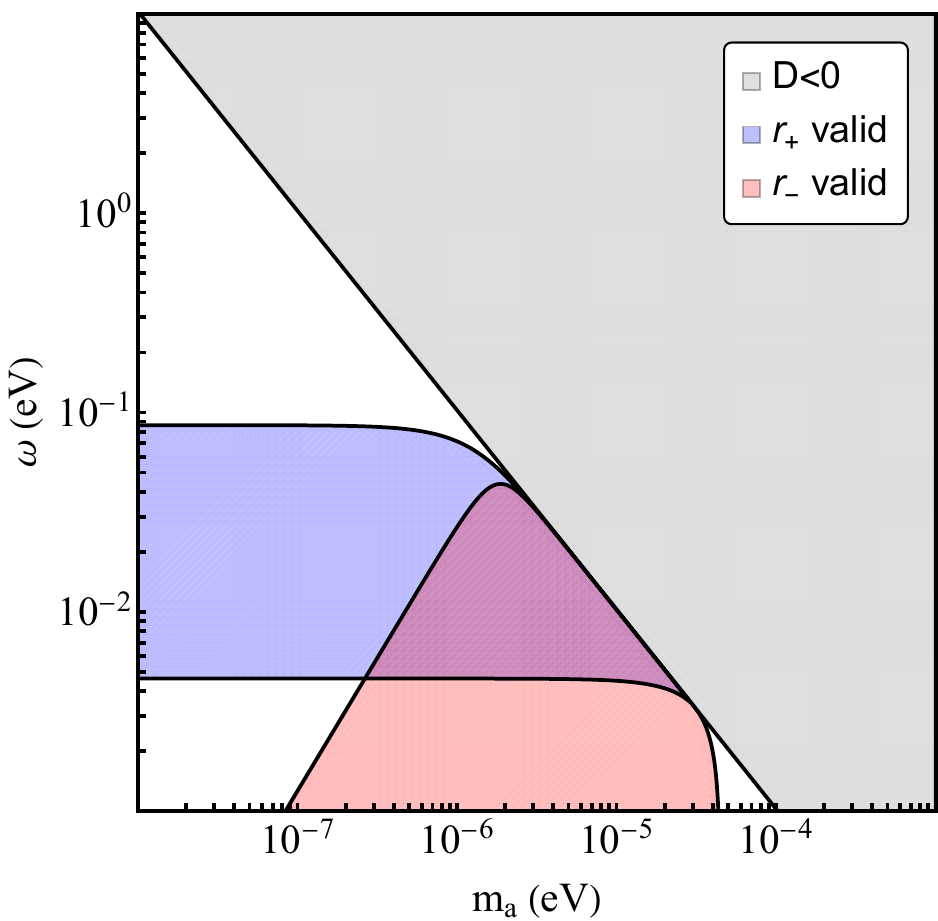}
    \\[2pt] {\small (a) 4U 0142+61 with $\kappa=1$.}
        \label{fig:4uk1}
    \end{minipage}\hfil
    \begin{minipage}{0.4\linewidth}
        \centering
        \includegraphics[width=8 cm]{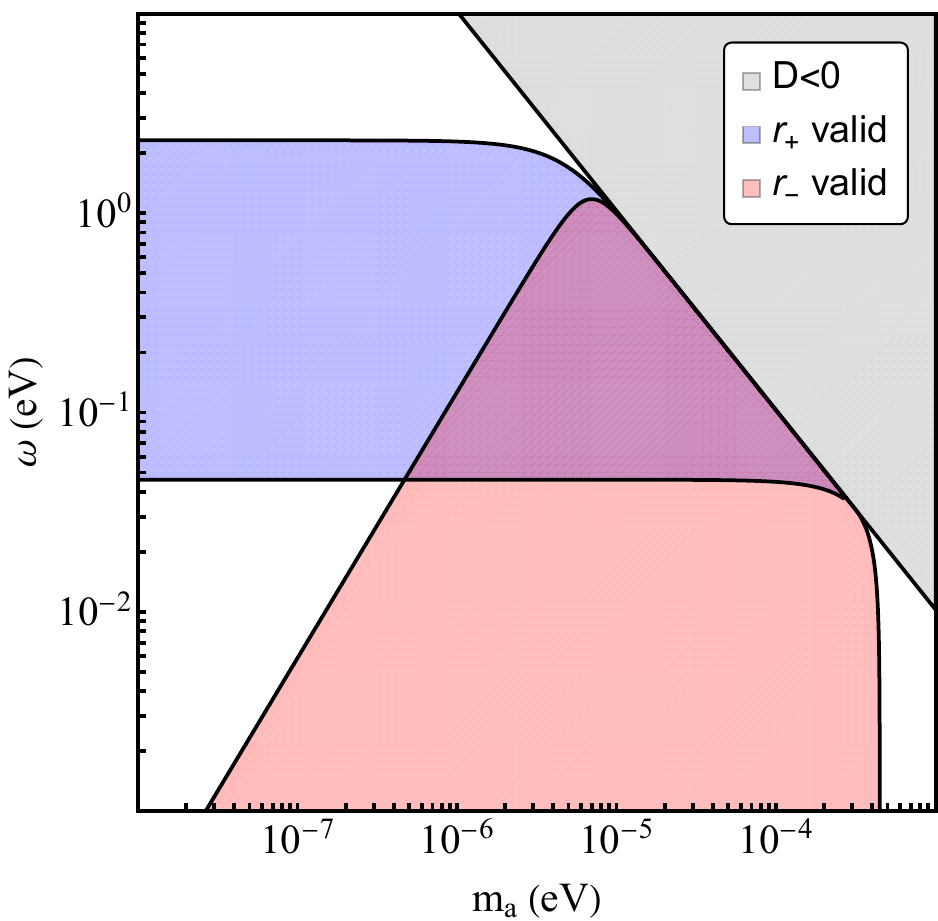}
        \\[2pt] {\small (b) 4U 0142+61 with $\kappa=100$.}
        \label{fig:4uk100}
    \end{minipage}

    \vspace{0.4cm}

    \begin{minipage}{0.4\linewidth}
        \centering
        \includegraphics[width=8 cm]{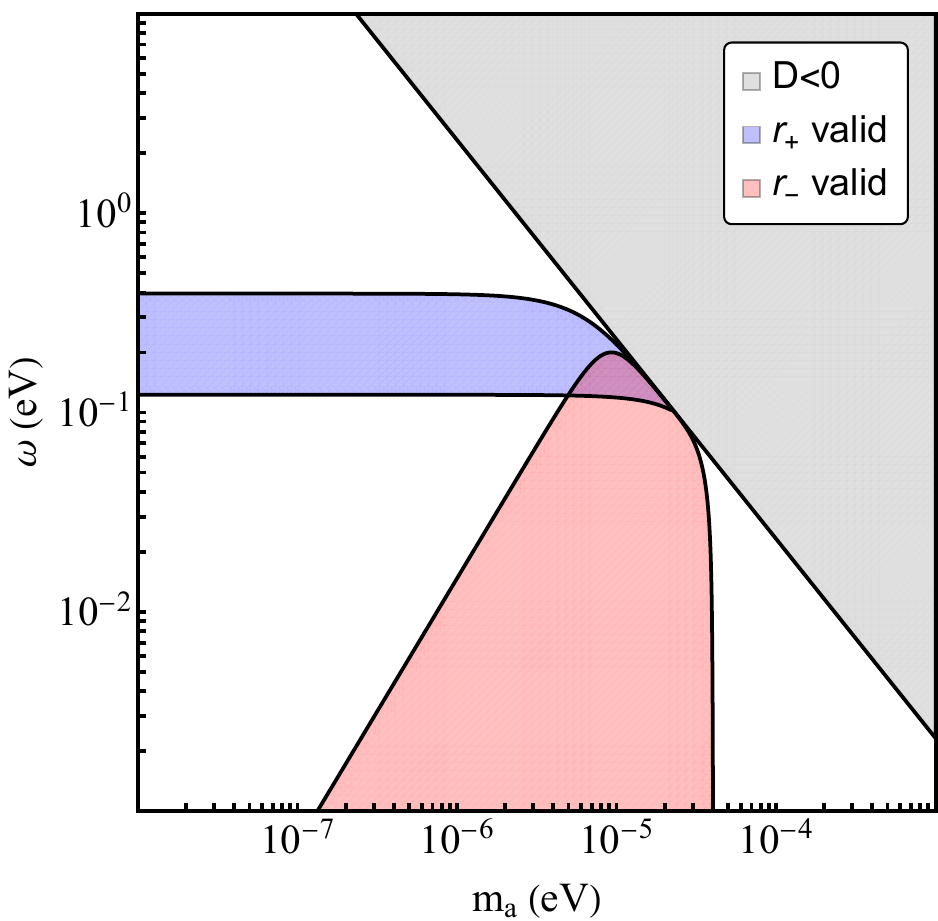}
        \\[2pt] {\small (c) PSR B0656+14 with $\kappa=1$.}
        \label{fig:b0656k1}
    \end{minipage}\hfil
    \begin{minipage}{0.4\linewidth}
        \centering
        \includegraphics[width=8 cm]{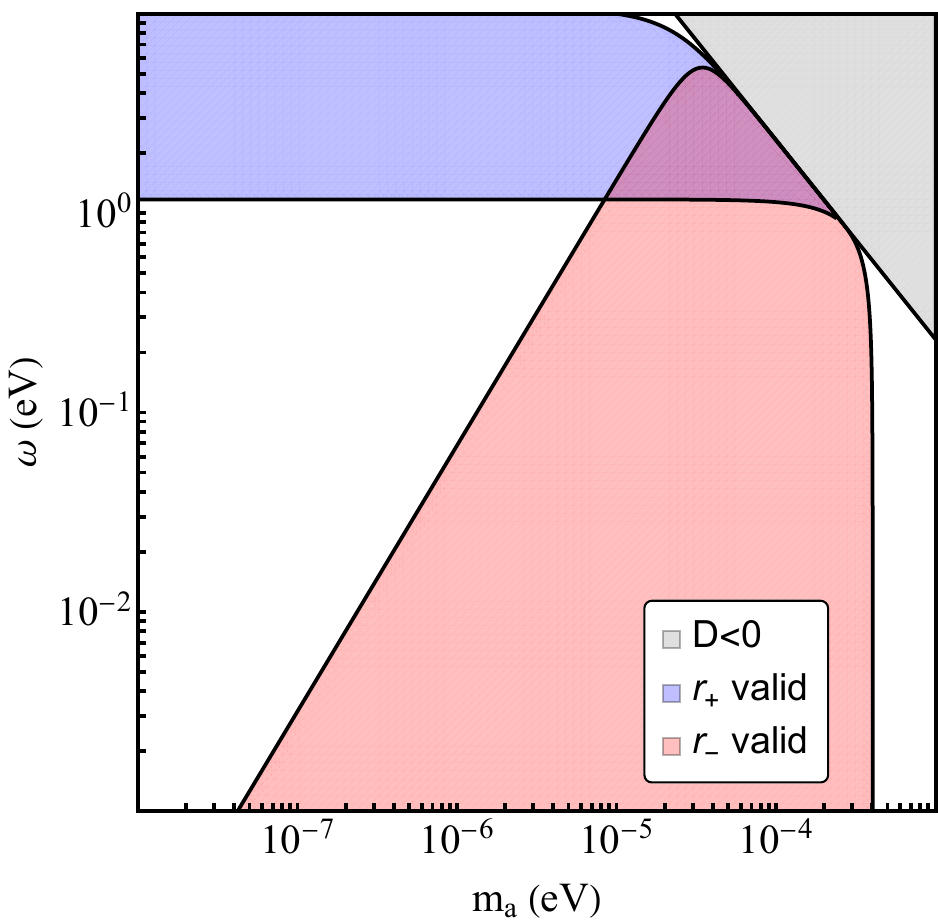}
        \\[2pt] {\small (d) PSR B0656+14 with $\kappa=100$.}
        \label{fig:b0656k100}
    \end{minipage}

    \caption{Validity regime of the LZ formula for the two stars 4U 0142+61 (upper row) and PSR B0656+14 (lower row) for two benchmark values of pair multiplicity, $\kappa=1$ (left column) and $\kappa=100$ (right column).  We have set $\Omega t=0$.}
    \label{fig:lzvalidity}
\end{figure*}

For fixed stellar parameters, the validity of the LZ approximation is determined by the axion mass $m_a$, the photon energy $\omega$, and the plasma multiplicity $\kappa$. 
Fig.~\ref{fig:lzvalidity} summarizes the regions of parameter space where the LZ approximation is expected to be applicable. The gray region denotes the non-resonant regime given by Eq.~\eqref{eq:masscondition}. The blue and red regions indicate where the vacuum- and plasma-like resonances, respectively, satisfy the localization criteria described in Sec.~\ref{sec:lzvalcons}. Outside these regions, although formal resonance solutions to Eq.~\eqref{sextic} may exist, the resonance either overlaps with the stellar surface or becomes too broad compared to the background variation scale. Specifically, above (left of) the blue (red) area the width of the resonance becomes too large. Below (right of) the blue (red) area the resonance overlaps with and eventually moves inside the star. The overlapping regime of validity further requires that the resonances can be independently analyzed, ie. that 
Eq.~\eqref{eq:nooverlapcondition} holds, and we find this to be the case for these benchmarks. As the simultaneous validity of LZ treatment for both resonances is rare, occurring only in a small overlap region, the validity of constraints derived from the LZ treatment depends heavily on whether the treatment is valid for the dominant resonance. We find that the vacuum-like resonance is typically dominant whenever it is present and obeys the localization criteria outlined above.

\begin{figure*}[t]
    \centering
    \begin{minipage}{0.48\linewidth}
        \centering
        \includegraphics[width=0.85\linewidth]{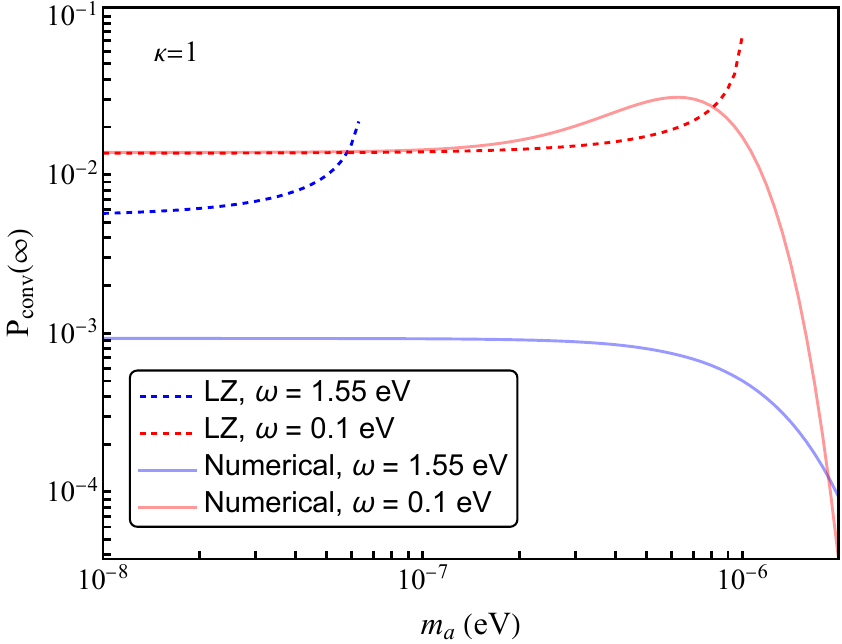}
    \end{minipage}\hfill
    \begin{minipage}{0.48\linewidth}
        \centering
        \includegraphics[width=0.85\linewidth]{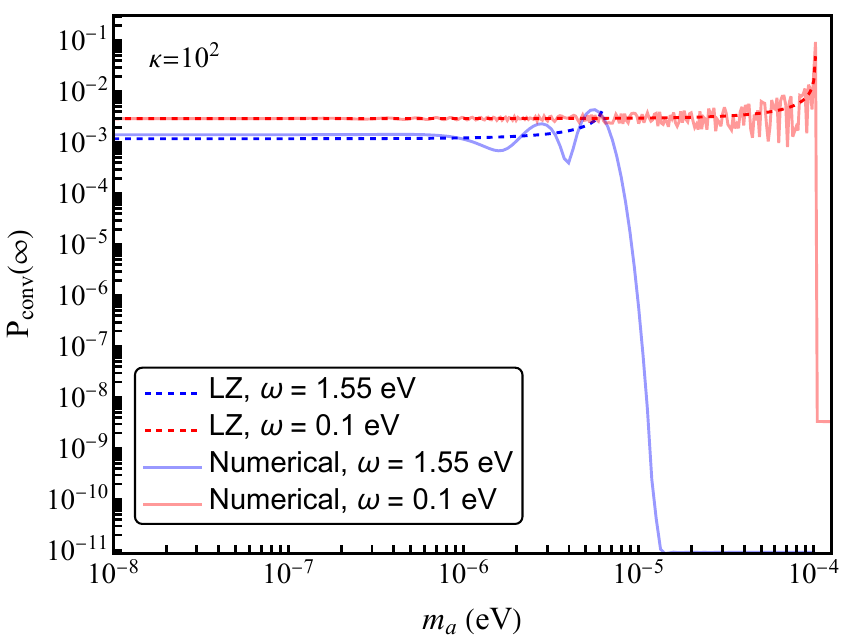}
    \end{minipage}
    \caption{Asymptotic conversion probability for two values of the photon frequency, $\omega=1.55$ eV and $\omega=0.1$ eV, and for two values of plasma multiplicity, $\kappa=1,10^2$. The neutron star parameters used correspond to magnetar 4U 0142+61, see Table~\ref{tab:neutronstars}. It is seen, that for $\kappa,\omega$ inside validity regimes of Fig.~\ref{fig:lzvalidity}, the LZ result tracks the numerical, and overestimates conversion otherwise. In this figure we set $g_{a\gamma}= 10^{-21}\mathrm{eV}^{-1}$ and $\Omega t=0$. }
    \label{fig:probvsma}
\end{figure*}

\begin{figure*}[t]
    \begin{minipage}{0.48\linewidth}
        \centering
        \includegraphics[width=0.85\linewidth]{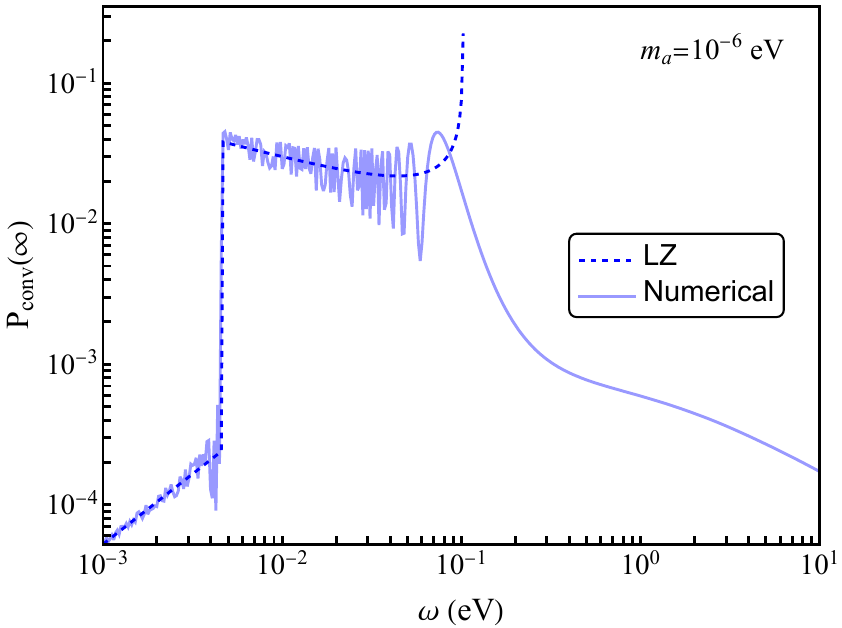}
    \end{minipage}\hfill
    \begin{minipage}{0.48\linewidth}
        \centering
        \includegraphics[width=0.85\linewidth]{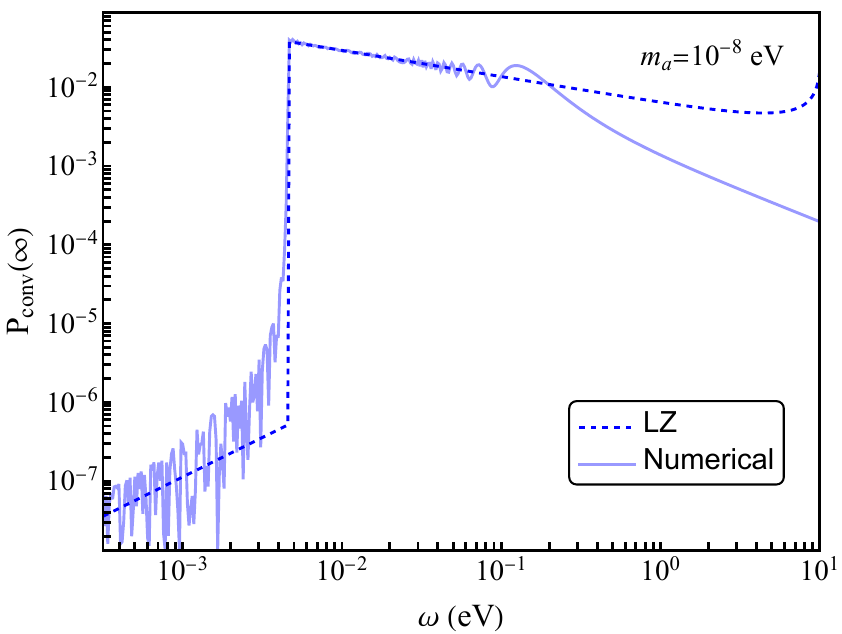}
    \end{minipage}
    
    \caption{Asymptotic conversion probability for two values of the axion mass, $m_a=10^{-6}$ eV and $m_a=10^{-8}$ eV. The neutron star parameters used correspond to magnetar 4U 0142+61, see Table~\ref{tab:neutronstars}. It is seen that for lower masses, the LZ result deviates from the numerical for intermediate values of $\omega$, corresponding to the gap between validity areas seen in Fig.~\ref{fig:lzvalidity}. We have set $g_{a\gamma}= 10^{-21}\mathrm{eV}^{-1}$ and $\Omega t=0$.}
    \label{fig:PvsOmega}
\end{figure*}
The upper panels in Fig.~\ref{fig:lzvalidity} correspond to magnetar 4U 0142+61 while the lower panels show PSR B0656+14. The left and right panels assume pair multiplicities $\kappa=1$ and $\kappa=100$, respectively. Increasing $\kappa$ shifts the vacuum resonance validity region to higher $\omega$ and broadens validity area of both the resonances.
From Fig.~\ref{fig:lzvalidity} we may immediately conclude that the validity of the LZ result in the optical regime is not generic and depends sensitively on various neutron star parameters, most importantly on the a priori undetermined plasma multiplicity $\kappa$.

To test the validity criteria derived earlier, we compare the asymptotic value of the conversion probability in Eq.~\eqref{Pconvr} to the result obtained from direct numerical integration of the Schrödinger equation of Eq.~\eqref{eq:schrödinger}. The results are shown as a function of $m_a$ for two fixed values of $\omega$ in Fig.~\ref{fig:probvsma} and as a function of $\omega$ for two fixed values of $m_a$ in Fig.~\ref{fig:PvsOmega}. In Fig.~\ref{fig:probvsma} we have varied the pair multiplicity $\kappa=1,10^2$ to see how it affects the conclusions. The numerical results are given by solid lines, while the LZ approximation tracks the dashed lines. For the LZ lines we have evaluated Eq.~\eqref{eq:LZformulabasic} with the general adiabaticities of Eq.~\eqref{generalgamma}. This requires that the resonances are well separated, which is the case for the parameter values considered in these figures.

Fig.~\ref{fig:probvsma} confirms the results shown in Fig.~\ref{fig:lzvalidity}. For $\kappa=1$ the numerical and LZ results are in agreement at $\omega=0.1 $ eV where the vacuum resonance satisfies the localization criteria. At the optical frequency $\omega=1.55$ eV, however, the LZ result substantially overestimates the conversion probability as the resonance width exceeds the scale over which the background is approximately constant. For increased plasma multiplicity, $\kappa=100$, the resonance is shifted back into the LZ validity regime of Fig.~\ref{fig:lzvalidity}, restoring agreement between the numerical and analytical results.

Fig.~\ref{fig:PvsOmega} provides the complementary comparison as function of the photon energy $\omega$. At low photon energies the conversion is dominated by the plasma-like resonance and at high energies by the vacuum-like resonance. For axion mass $m_a=10^{-6}$ eV both of the resonances are in the LZ validity regime and the numerical result tracks the LZ estimate very accurately until the the resonance vanishes according to Eq.~\eqref{eq:masscondition}. The low frequency scaling $P_\mathrm{conv}(\infty)\propto \omega$ and high frequency scaling $P_\mathrm{conv}(\infty)\propto \omega^{-1/3}$ match exactly those predicted by the LZ adiabaticity formulae of Eq.~\eqref{gammas}. For lower axion masses, here $m_a=10^{-8}$ eV, the low frequency numerical result begins to deviate from the LZ prediction, corresponding to the area between the plasma and vacuum validity areas in Fig.~\ref{fig:lzvalidity}. For higher frequencies one enters the vacuum resonance dominated regime, where the LZ approximation also holds, restoring agreement between the two results.

Taken together, Figs.~\ref{fig:PvsOmega} and~\ref{fig:probvsma} demonstrate that the agreement between the LZ approximation and the full numerical solution is controlled by the localization criteria given in Sec~\ref{sec:lzvalcons}. Whenever these conditions are satisfied, the LZ approximation accurately reproduces the numerical conversion probability. Conversely, when the resonance overlaps the stellar surface or becomes too broad compared to the background scale, the approximation systematically fails, typically overestimating the conversion probability in the optical regime relevant for instance for polarization searches.

\section{Polarization and axions}\label{sec:polarization}

\subsection{Linear degree of polarization}

We will now apply the general results of Sec.~\ref{sec:resonances} to describe the effect induced by resonant photon-axion conversion on photon polarization. We consider initially unpolarized thermal emission from the stellar surface. As only the $A_\parallel$-polarization mixes with axions, the outgoing radiation becomes linearly polarized in the $A_\perp$ direction as it traverses through the resonance.

The linear degree of polarization is defined as
\begin{align}
    \Pi_\mathrm{L}(r)
    \equiv
    \frac{\sqrt{Q^2+U^2}}{I}(r)
\end{align}
in terms of the Stokes parameters, $I$, $Q$ and $U$, which are defined as
\begin{align}
    I&=\langle|A_\parallel|^2\rangle_\mathrm{E}
      +\langle|A_\perp|^2\rangle_\mathrm{E},
      \notag\\
    Q&=\langle|A_\parallel|^2\rangle_\mathrm{E}
      -\langle|A_\perp|^2\rangle_\mathrm{E},
      \notag\\
    U&=
    2\,\mathrm{Re}
    \!\left(
    \langle A_\parallel A_\perp^\ast\rangle_\mathrm{E}
    \right). \label{eq:stokes}
\end{align}
For the average over an ensemble of random phases, denoted by $\langle\cdot\rangle_{\text{E}}$ in the above definitions, we have $\langle |A_\||^2\rangle_{\text{E}}=|A_\||^2$ and $\langle |A_\perp|^2\rangle_{\text{E}}=|A_\perp|^2$. Furthermore, we choose the polarization basis such that $U=0$ initially, which remains true as the basis rotation is slow. Using these results, we obtain
\begin{align}
\Pi_\mathrm{L}(r)
\approx
    \frac{\big||A_\parallel|^2-|A_\perp|^2\big|}
    {|A_\parallel|^2+|A_\perp|^2}.
    \label{linpol}
\end{align}

For initially unpolarized radiation,
\begin{align}
    |A_\parallel(R)|^2
    =
    |A_\perp(R)|^2
    =
    \frac12 .
    \label{eq:polarizationinicond}
\end{align}
Since only the parallel mode mixes with the axion,
\begin{align}
    |A_\parallel(r)|^2
    =
    P_\mathrm{S}(r)\,
    |A_\parallel(R)|^2,
    \qquad
    |A_\perp(r)|^2
    =
    |A_\perp(R)|^2 ,
\end{align}
where $P_\mathrm{S}$ denotes the photon survival probability. Substituting these expressions into Eq.~\eqref{linpol} gives
\begin{align}
    \Pi_\mathrm{L}(r)
    =
    \frac{1-P_\mathrm{S}(r)}
         {1+P_\mathrm{S}(r)}.
    \label{eq:PiLgeneral}
\end{align}
Using the stationary-phase approximation from Eq.~\eqref{eq:aparallorder} yields
\begin{align}
    P_\mathrm{S}(\infty)
    =\frac{|A_\|(\infty)|^2}{|A_\|(R)|^2}=
    \cos^2
    \!\left(
        \sqrt{2\pi\gamma_\mathrm{res}}
    \right),
\end{align}
and therefore
\begin{align}
    \Pi_\mathrm{L}(\infty)
    =
    \frac{
        \sin^2\!\left(
            \sqrt{2\pi\gamma_\mathrm{res}}
        \right)
    }{
        1+
        \cos^2\!\left(
            \sqrt{2\pi\gamma_\mathrm{res}}
        \right)
    }.
    \label{eq:PiLSPA}
\end{align}
Eq.~\eqref{eq:PiLSPA} constitutes the general stationary phase prediction for the linear degree of polarization. In the non-adiabatic regime,
\begin{align}
    \gamma_\mathrm{res}\ll 1,
\end{align}
the result reduces to
\begin{align}
    \Pi_\mathrm{L}(\infty)
    =
    \pi\gamma_\mathrm{res}
    +
    \mathcal O(\gamma_\mathrm{res}^3)
    \simeq -\Re \Big( A_{\|}^{(2)}(\infty )\Big),
    \label{eq:lindegfinal}
\end{align}
where in the last equality we used Eq.~\eqref{eq:amp2ndorder} to write the polarization in terms of the leading perturbation theory contribution to the photon survival amplitude. 

\begin{figure}
\includegraphics[width=7cm]{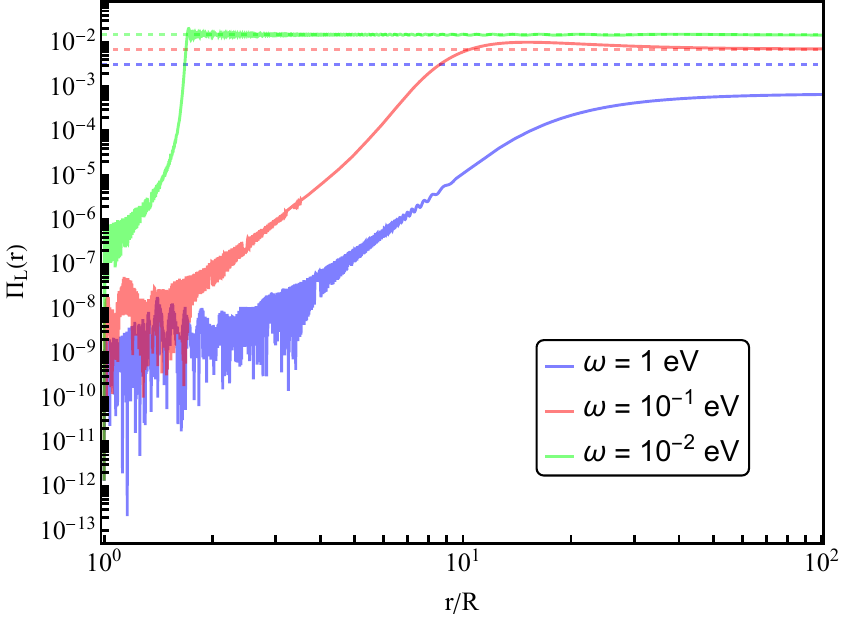}
\caption{Accumulation of linear polarization with distance due to the vacuum resonance. We have fixed $m_a=10^{-11}$ eV, $g_{a\gamma}=10^{-21} \,{\mathrm{eV}}^{-1}$, $\Omega t =0$, $\kappa=1$ and used the stellar parameters of magnetar 4U 0142+61 shown in Table~\ref{tab:neutronstars}. The dotted lines correspond to the LZ formula.}
\label{fig:polwithdist}
\end{figure}

Fig.~\ref{fig:polwithdist} shows how the polarization evolves with distance, in the case of very low axion mass, $m_a=10^{-11}$~eV. We have solved the Schrödinger equation numerically, and 
the results are shown by solid lines in the figure. The LZ predictions are shown with dotted lines, and the figure shows how the numerical result approaches the asymptotic result of the LZ formula at large distances. For small conversion probabilities and when the LZ formula holds, we have $\Pi_\mathrm{L}(\infty)=\pi \gamma_\mathrm{res}$ and the asymptotic polarization from the numerical computation follows the LZ prediction. For higher photon frequencies the width of the vacuum resonance increases, and LZ formula begins to overestimate the conversion strength. We can see that in the case of optical light emitted from the magnetar 4U 0142+61 with $\kappa=1$, the LZ formula is not applicable. However, due to uncertainty in the value of $\kappa$, this conclusion may be altered for e.g. $\kappa=100$; see the top right panel in Fig.~\ref{fig:lzvalidity}. 

Finally, we remark on the effect of stellar rotation on the polarization signal. Strictly speaking, the Stokes parameters are also functions of $\Omega t$. Thus, in the case of non-phase-resolved polarimetry the ensemble averages have to be further averaged over rotation. This leads to Eq.~\eqref{eq:PiLgeneral} being given instead with respect to the averaged photon survival probability $\langle P_\mathrm{S}(r)\rangle_{\Omega t}$. For the small conversion probabilities considered here, this is numerically equivalent to simply averaging the linear polarization degree $\Pi_\mathrm{L}$ over the rotation.

\subsection{Polarization constraints on axions}
We now consider constraints on axions from existing linear polarization data. While optical frequency data is currently sparse, existing 8-10 meter class optical and near-IR telescopes such as the Very Large Telescope (VLT), Subaru, and the Southern African Large Telescope (SALT) are all equipped with optical polarimeters~\cite{cikota2016linear,2002PASJ...54..819K,salt}, with the first two having already been used for optical frequency polarization measurements. Here we consider constraints from polarization measurements of the rotation-powered pulsar PSR B0656+14 (VLT)~\cite{Mignani:2015ufa} and the magnetar 4U 0142+61 (Subaru)~\cite{Wang:2015ppa}. Additionally, we consider a hypothetical optical benchmark scenario, with parameters $\Pi_\mathrm{L}<1 \% $, $R=12$ km, $B_0=10^{14}$ G, $T=10 $ s, $\omega=1$ eV. We note that next generation telescopes with polarimeters operating in the mid-IR frequencies are expected to provide even greater sensitivity.

\begin{figure*}[hbt!]
    \begin{minipage}{0.32\linewidth}
        \centering
        \includegraphics[width=.9\linewidth]{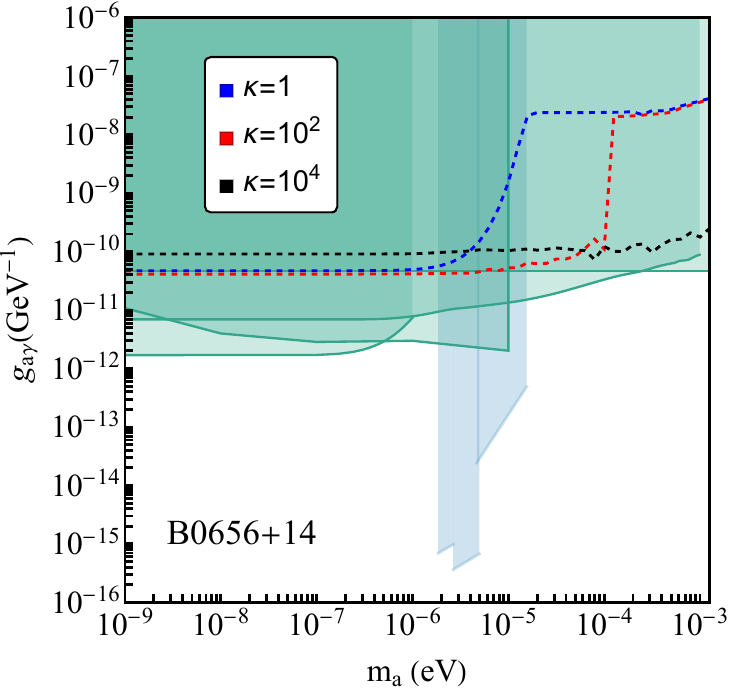}
    \end{minipage}\hfill
    \begin{minipage}{0.32\linewidth}
        \centering
        \includegraphics[width=.9\linewidth]{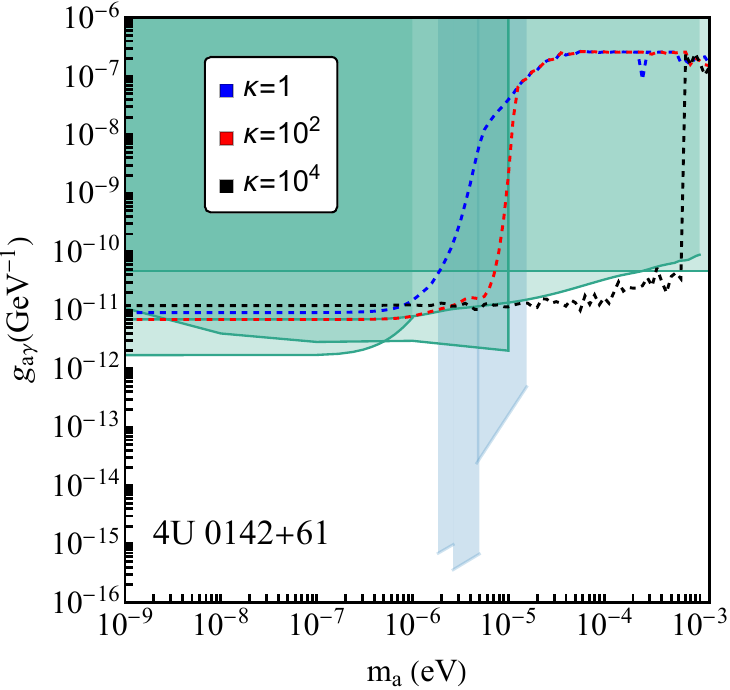}
    \end{minipage}\hfill
    \begin{minipage}{0.32\linewidth}
        \centering
        \includegraphics[width=.9\linewidth]{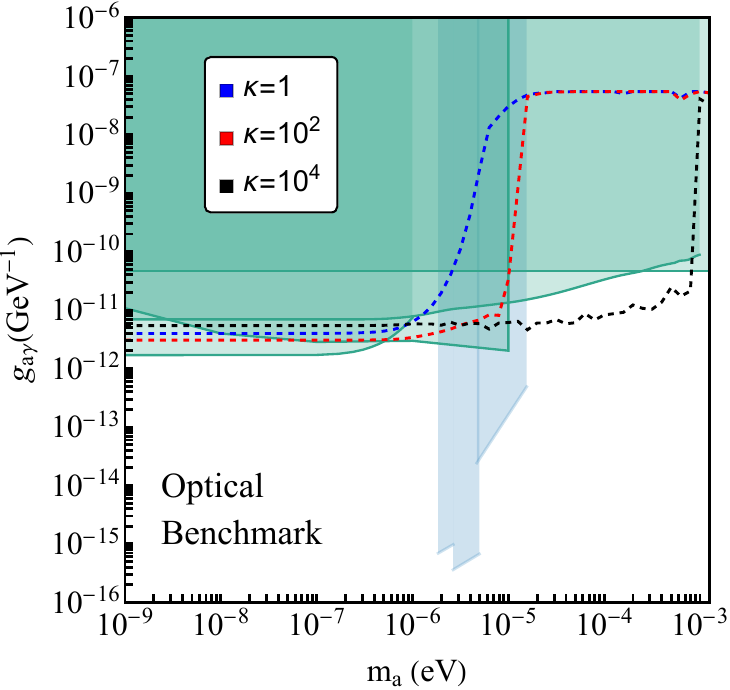}
    \end{minipage}
    \caption{Sensitivities of linear polarization optical searches from PSR B0656+14, magnetar 4U 0142+61 and an optical benchmark respectively under varying assumptions of plasma multiplicity $\kappa$. The existing constraints in various shades of green shown are from ADMX~\cite{ADMX:2018gho,ADMX:2019uok}, RBF+UF~\cite{Wuensch:1989sa,Hagmann:1996qd}, magnetic white dwarf polarization~\cite{Benabou:2025jcv}, pulsar polar caps~\cite{Noordhuis:2022ljw} and Globular Clusters~\cite{Dolan:2022kul}.}
    \label{fig:constraints}
\end{figure*}

For each star we numerically solve the Schrödinger equation and compute the linear degree of polarization using the results of Sec.~\ref{sec:polarization}. To account for the rotation of the star, we have averaged the linear polarization degree over 50 evenly spaced values of the rotation phase $\Omega t \subset [0,\pi]$. Constraints are then obtained by finding the coupling which saturates the upper bound of the inequality $\langle \Pi_\mathrm{L}(\infty)\rangle_{\Omega t}<\Pi_\mathrm{obs}$, where the observed polarization is given in Table~\ref{tab:neutronstars}.  
The resulting sensitivities are shown in Fig.~\ref{fig:constraints}, where we show constraints for benchmark values of the plasma multiplicity $\kappa=1,10^2,10^4$.

The interpretation of these constraints is, however, subject to several astrophysical uncertainties. First, the dipolar magnetosphere adopted throughout this work may not accurately describe the magnetic field close to magnetars, where higher-order multipoles and twisted field configurations are expected. These effects are particularly relevant for the plasma resonance, which approaches the stellar surface for increasing axion mass~\cite{Roy:2025mqw}. Second, the plasma pair multiplicity $\kappa$ is poorly constrained observationally. In ordinary pulsars, it is typically modest, $\kappa \sim \mathcal{O}(1)$, whereas in magnetars copious amounts of electron-positron pairs can be produced, leading to much larger multiplicities of $\kappa \sim 10^2$--$10^4$~\cite{Timokhin:2018vdn}. Since $\kappa$ is not directly observable, we have presented results for several benchmark values.
Third, radiative transfer effects such as cyclotron scattering in strong magnetic fields generally lead to much larger intrinsic polarizations~\cite{Nobili:2008qw}. This generally weakens the axion interpretation of the signal. Finally, optical and infrared emission from magnetars may contain significant non-thermal components originating from the magnetosphere, including surface hotspots and outer magnetospheric currents~\cite{vanAdelsberg:2006uu}. Emission from higher altitudes may avoid the resonance, or traverse through only a part of it. This is especially important for the vacuum resonance, which moves towards the star and narrows as $\kappa$ is increased. The no-overlap condition may be generalized to $r_\mathrm{res}-\frac{1}{2}L_\mathrm{res}\gtrsim r_\mathrm{em}$ where $r_\mathrm{em}$ is the emission radius. Therefore, for the same star, the modified no-overlap condition is harder to satisfy and the LZ approximation is more likely to fail. Specifically, three cases can be distinguished. If $r_{\rm em}\lesssim r_{\rm res}-L_{\rm res}/2$, the resonance is fully traversed, and both the LZ formula and the full result are independent of $r_{\rm em}$. If $r_{\rm em}$ lies within the resonance width, the conversion region is truncated and the LZ formula, which depends only on quantities at $r_{\rm res}$, does not capture the resulting reduction. If $r_{\rm em}$ lies beyond the resonance, it is not traversed at all and only non-resonant conversion remains. The LZ formula then correctly predicts no resonant contribution, and the full calculation includes the non-resonant part automatically. For large $\kappa$ the vacuum resonance is narrow and close to the star. The second case then occupies only a thin range of $r_{\rm em}$, and the LZ treatment remains reliable, while the conversion itself depends strongly on whether the emission point lies below or above the resonance. For small $\kappa$ the resonance is so broad that the localized level-crossing picture no longer holds for any emission radius, including the stellar surface. The conversion is then a coherent build-up over an extended region where the phase is nearly frozen, and it is only weakly affected by the emission altitude as long as the emission point lies below the region where the conversion builds up. In all cases emission from higher altitudes weakens the conversion, and hence the constraints, and the full calculation accounts for this directly. Our conclusions regarding the validity of the LZ approximation therefore persist for emission above the stellar surface. The effects of magnetospheric emission in our context were also considered in Ref.~\cite{Song:2024rru}

These uncertainties are especially important for the constraints from magnetar 4U 0142+61 in Fig.~\ref{fig:constraints}. For pulsars the constraints are weaker, but the uncertainties are substantially alleviated.

Despite these caveats, Fig.~\ref{fig:constraints} demonstrates that optical polarization measurements provide sensitivity complementary to existing laboratory and astrophysical searches. More importantly, our results show that reliable interpretation of such observations requires a numerical treatment of axion-photon propagation whenever the localization criteria derived in Sec.~\ref{sec:lzvalcons} are not satisfied. In particular, for optical and near-infrared observations the LZ approximation frequently overestimates the conversion probability, leading to overly optimistic constraints if used outside its regime of validity.
 
\begin{figure*}[hbt!]
    \centering
    \begin{minipage}{0.32\linewidth}
        \centering
        \includegraphics[width=.9\linewidth]{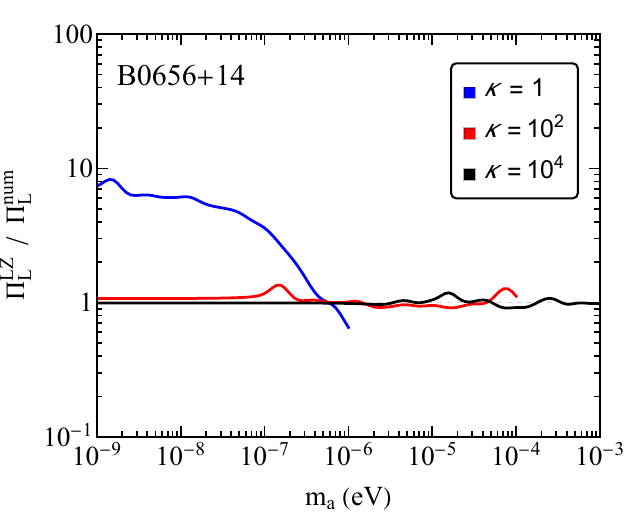}
    \end{minipage}\hfill
    \begin{minipage}{0.32\linewidth}
        \centering
        \includegraphics[width=.9\linewidth]{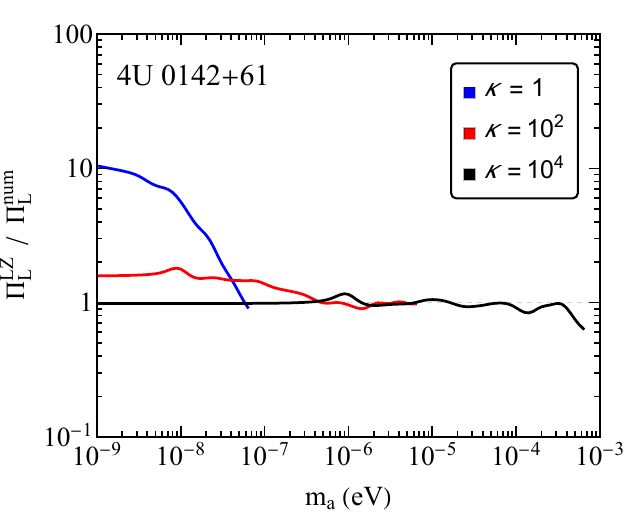}
    \end{minipage}\hfill
    \begin{minipage}{0.32\linewidth}
        \centering
        \includegraphics[width=.9\linewidth]{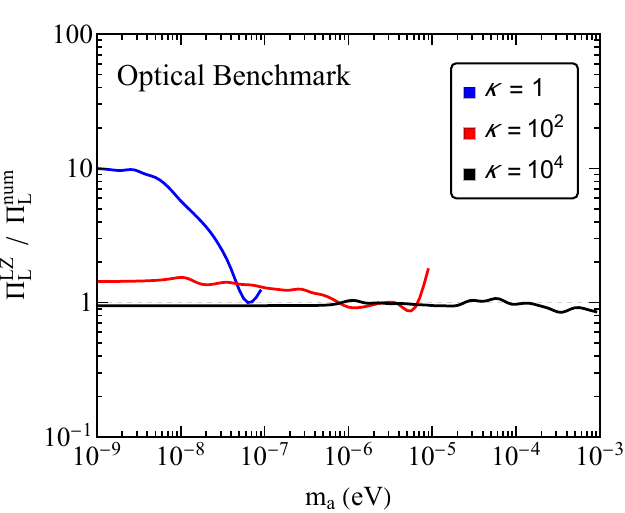}
    \end{minipage}
    \caption{Ratio of the LZ prediction of phase-averaged linear polarization and the numerical result. As the polarization depends on the square of the coupling constant $g_{a\gamma}$, the effect on the sensitivity shown in Fig. \ref{fig:constraints} is given by the square root of this ratio.}
    \label{fig:constraints2}
\end{figure*}

Fig.~\ref{fig:constraints2} shows the phase-averaged polarization signal, as predicted by the LZ formula, and the fully numerical calculation. The results are consistent with Figs.~\ref{fig:probvsma}, \ref{fig:PvsOmega}: LZ overestimates conversion for small $\kappa$, by a factor of approximately $10$, which shifts the constraints by its square root. The LZ curves terminate abruptly when the two resonances merge.

\section{Conclusions and outlook}\label{sec:conclusions}

Recently, it has been shown that light emanated from neutron star magnetospheres in sub-millimeter to optical ($\omega\sim 10^{-3}-1$ eV) band  
may convert resonantly to axions
producing sharp spectral features and polarization signals~\cite{Song:2024rru,Bondarenko:2022ngb}. These analyses have relied heavily on the LZ approximation.

Thus, we first established the connection between the LZ formula and the all-orders perturbative stationary phase treatment. We then used the stationary phase result to provide simple criteria determining whether the resonances remain sufficiently localized for the LZ approximation to apply. Our main result is that for photon frequencies where the axion conversion would be expected to proceed via vacuum resonance, the assumptions underlying the LZ formula are frequently violated. We showed how the plasma pair multiplicity acts as the most important astrophysical nuisance parameter affecting this result. 
Our analysis was based on localization criteria for the resonances, and we 
compared this against direct numerical integration. Numerical computation 
agrees with the analytic computation based on localization criteria and shows 
that when they are invalid, the LZ formula systematically overestimates the conversion probability.

As an application of these results we considered constraints from axion-induced polarization in neutron star thermal emission. We find the constraints to be slightly weaker than those obtained in earlier literature using the LZ approximation~\cite{Song:2024rru}.

Nevertheless, future millimeter-to-optical band observations remain a promising avenue for axion detection across a wide range of masses $m_a\lesssim 10^{-5}-10^{-3}$ eV, depending on plasma multiplicity. While we have illustrated the impact of the breakdown of the LZ approximation on axion polarization constraints from neutron stars, our conclusions are more general. They apply equally to all observables relying on resonant axion-photon conversion, such as spectral features and flux suppression. Robust constraints from these phenomena therefore require a more accurate picture of the resonant conversion than given by the LZ formula.

\begin{acknowledgments}
This work has been supported by the Research Council of Finland (grants\# 342777, 371542). TS would like to thank the Vilho, Yrjö and Kalle Väisälä Foundation. 
\end{acknowledgments}


\appendix

\section{Magnetospheric Model and Photon Propagation}
\label{app:magnetosphere}

Here we describe in more detail our assumptions of the neutron star magnetosphere. We assume the magnetic field to be accurately described by a misaligned dipole rotator. In this case the magnetic field along a vector $\hat{\mathbf{r}}$ pointing towards an observer is given by
\begin{align}
    \mathbf{B}(\mathbf{r})=B_0\left(\frac{R}{r}\right)^3\Big(3 (\hat{\mathbf{m}} \cdot \hat{\mathbf{r}})\hat{\mathbf{r}}-\hat{\mathbf{m}}\Big),
\end{align}
with $B_0$ the surface magnetic field strength and R the stellar radius, and with the misalignment angle between the rotation and magnetic axes governed by
\begin{align}
    \hat{\mathbf{m}} \cdot \hat{\mathbf{r}}=\cos \theta_\mathrm{m} \cos \theta_{\mathrm{o}}+\sin \theta_\mathrm{m} \sin \theta_{\mathrm{o}} \cos (\Omega t), \label{mr}
\end{align}
with $\theta_\mathrm{m},\theta_\mathrm{o}$ the polar angles from the rotation axis $\hat{\bf{z}}$ to the magnetic and observation angle respectively, and $\Omega t$ the phase angle of observation.
The magnitude of the magnetic field is given by 
\begin{align}
    B({\bf{r}})&=({\bf{B}}({\bf{r}})\cdot {\bf{B}}({\bf{r}}))^{1/2} \notag \\ 
   &= B_0 \left(\frac{R}{r}\right)^3\sqrt{3(\hat{\mathbf{m}} \cdot \hat{\mathbf{r}})^2+1}\notag \\
   &\equiv B_0 \left(\frac{R}{r}\right)^3F(\theta_\mathrm{m},\theta_\mathrm{o},\Omega t) .
\end{align}

To model the plasma, we use the Goldreich-Julian (GJ) model~\cite{GJ}, but allow the charge density to be multiplied by a pair multiplicity factor $\kappa$ which encodes the likely enhancement of the plasma density due to pair creation in strong fields. This factor is typically in the range $\kappa\sim 1-10^4$~\cite{Timokhin:2018vdn}. The charge density is in general given by
\begin{align}
    \rho_e(\mathbf{r}) &=\kappa\times2 \boldsymbol{\Omega} \cdot \mathbf{B} \frac{1}{1-\Omega^2 r^2 \sin ^2 \theta_{\mathrm{0}}}\notag \\ 
    &\simeq \kappa \times \frac{4\pi}{T}B_z, \label{gjne}
\end{align}
where the angular momentum of the star is aligned as $\boldsymbol{\Omega}=\Omega \hat{\mathbf{z}}=\frac{2\pi}{T}\hat{\mathbf{z}}$ with $T$ the orbital period. 
The relativistic correction factor is not relevant for the slow rotators relevant to polarization signals considered in this paper. The magnetic field projection along the rotation axis is given by
\begin{align}
    B_z&=B_0\left(\frac{R}{r}\right)^3\Big(3 \cos \theta_\mathrm{o} (\hat{\mathbf{m}} \cdot \hat{\mathbf{r}})-\cos \theta_\mathrm{m}\Big)\notag \\
    &\equiv B_0\left(\frac{R}{r}\right)^3G(\theta_\mathrm{m},\theta_\mathrm{o},\Omega t).
\end{align}

The number density is then $n_e=|\rho_e(\mathbf{r})|/e$. Assuming that electrons dominate the charge density of the magnetosphere, the plasma frequency is then given as
\begin{align}
    \omega_\mathrm{pl}(\mathbf{r})&=\sqrt{\frac{4\pi\alpha n_e(\mathbf{r})}{m_e}}.
\end{align}

Next, to facilitate analytical computations, we must make some assumptions regarding photon propagation. Specifically, the angle $\theta$ between the photon momentum and magnetic field governs the strength of the axion-photon mixing, and we want to express it in terms of the observable angles $\theta_\mathrm{m},\theta_\mathrm{o}$. To do this, we assume that the photons move radially, $\vec{\mathbf{k}}=k\hat{\mathbf{r}}$. 
In this case one finds
\begin{align}
    \theta(\mathbf{r})&=\arccos \Bigg[\frac{2(\hat{\mathbf{m}} \cdot \hat{\mathbf{r}})}{3(\hat{\mathbf{m}} \cdot \hat{\mathbf{r}})^2+1}\Bigg],
\end{align}
with $\hat{\mathbf{m}} \cdot \hat{\mathbf{r}}$ given by Eq.~\eqref{mr}. We find that this angle depends somewhat sensitively on the magnetic and observational axis inclinations. 

\bibliographystyle{apsrev4-2}
\bibliography{apssamp}

\end{document}